\documentclass[12pt,a4paper]{article}
\usepackage{bbold}
\usepackage{graphicx}
\usepackage{soul}

\newcommand{\abs}[1]{\left|#1\right|}
\newcommand{\ket}[1]{\left|#1\right\rangle}
\newcommand{\tr}{\mathrm{tr}}
\newcommand{\A}{\mathcal{A}}

\makeatletter
 \def\@biblabel#1{#1.}
\makeatother
\begin{document}
\begin{center}
{\large \bf Metric Structure of the Space of Two-Qubit Gates, Perfect
  Entanglers and Quantum Control}\\
\bigskip
Paul Watts\footnote{{\tt watts@thphys.nuim.ie}, {\tt
    maurice.oconnor.2012@nuim.ie}, {\tt jiri.vala@nuim.ie}}${}^{,2}$,
Maurice O'Connor${}^1$ and Ji\v{r}\'i Vala${}^{1,2}$\\

\bigskip

${}^1$Department of Mathematical Physics,
National University of Ireland Maynooth,
Maynooth, Co.\ Kildare, Ireland\\
${}^2$School of Theoretical Physics,
Dublin Institute for Advanced Studies,\\
10 Burlington Road,
Dublin 4, Ireland

\bigskip

(Published in {\it Entropy} {\bf 15} (2013) 1963-1984)
\end{center}

\begin{abstract}
We derive expressions for the invariant length element and measure for
the simple compact Lie group $SU(4)$ in a coordinate system
particularly suitable for treating entanglement in quantum information
processing.  Using this metric, we compute the invariant volume of the
space of two-qubit perfect entanglers.  We find that this volume
corresponds to more than $84\%$ of the total invariant volume of the
space of two-qubit gates.  This same metric is also used to determine
the effective target sizes that selected gates will present in any
quantum-control procedure designed to implement them.
\end{abstract}

\noindent MSC Classification (2010): 81P68, 28C10, 22C05

\noindent

\setcounter{footnote}{0}
\section{Introduction}
\setcounter{equation}{0}
\renewcommand{\theequation}{\thesection.\arabic{equation}}

Unitary transformations of the states of two quantum bits (qubits)
play a prominent role in quantum information processing and
computation \cite{Kitaev}.  Physically, these quantum logic gates are
generated by interactions between qubits and thus the vast majority of
them are entangling operations, meaning that they can change the
degree to which the states of two qubits are strongly correlated or
entangled.  The entangling two-qubit operations, together with
suitable single-qubit gates, are also essential for universal quantum
computation.

Two-qubit operations are elements of the Lie group $SU(4)$ and so are
conveniently represented by $4\times 4$ unitary matrices of unit
determinant.  A comprehensive survey of such two-qubit gates is
offered by their geometric theory, which was formulated by Zhang et
al.\ \cite{ZVSW1}.  This uses both the Cartan decomposition of $SU(4)$
and the theory of local invariants of two-qubit operations
\cite{Makhlin} to provide a very useful geometric classification of
the two-qubit gates in terms of their local equivalence classes.
These classes are the two-qubit operations that are equivalent up to
single-qubit transformations, and thus each class is characterised by
its unique nonlocal content and thus its unique entangling
capabilities.  The geometric theory of two-qubit gates has recently
been utilised in the context of the physical generation of these gates
using an optimal-control approach \cite{MRMYVWCK}.

The geometric theory also provides a useful framework for the
characterisation of the specific two-qubit gates of most interest in
quantum computing.  These include not only familiar logical operations
like CNOT and SWAP, but also perfect entanglers, gates that are
capable of creating a maximally-entangled state out of some initial
product state.  Where these gates are located in $SU(4)$, and the
nature of the regions they are in, are issues that can only be
properly understood when the geometric structure of $SU(4)$ is
determined.

This geometry will have a major impact on the implementation of any
working quantum computer.  In constructing its gates, we need to know
where they are in $SU(4)$ and how likely it is that we can generate
them.  For instance, it was shown \cite{ZVSW1} that perfect entanglers
occupy exactly half of the volume of the space of all local
equivalence classes of two-qubit gates.  This naively suggests that if
one randomly picks a nonlocal gate, there will be a $50\%$ probability
that it is a perfect entangler.  This same picture also implies that
all gates are equally probable; picking a gate at random is just as
likely to produce a gate locally-equivalent to a CNOT gate as it is to
give one locally-equivalent to a SWAP.

However, this view ignores the local (i.e., single-qubit) operations
that are factored out from the local equivalence classes.  These
operations are represented by the $SU(2)\otimes SU(2)$ subgroup whose
curvature contributes to the overall geometry of $SU(4)$, and thus to
the distribution of locally-equivalent gates.  To incorporate this
curvature so as to correctly determine how the local equivalence
classes are distributed, we must find an invariant Haar measure for
$SU(4)$.

These considerations motivate the present work.  We first focus on the
derivation of the metric structure of $SU(4)$, specifically its
invariant length element and its Haar measure.  We would like to point
out that even though calculations using the Haar measure for various
Lie groups, including $SU(4)$, have been carried out in the past
\cite{MMM,TBS,SHH}, they were not performed in the representation
particularly applicable to dealing with entanglement in quantum
information processing, namely, one that reflects the natural
factorisation of $SU(4)$ into the single-qubit $SU(2)\otimes SU(2)$
and purely nonlocal (two-qubit) $SU(4)/SU(2)\otimes SU(2)$ parts.
This factorisation leads to a reduction from fifteen-dimensional
$SU(4)$ to a three-dimensional space in which all locally-equivalent
gates live, and we discuss the form of the length element and measure
for two particular choices of coordinates for this space.

We then use these derived geometric quantities to proceed towards our
main objective: the calculation of the invariant volumes of the
regions containing particular gates of interest in quantum information
processing.  First, we determine the total volume of the region
occupied by perfect entanglers, and find the rather surprising result
that these gates make up over $84\%$ of $SU(4)$ (thus quantifying the
statement that most of the two-qubit operations are perfect
entanglers).  We then consider regions containing the gates most often
used in quantum computing and find that their volume depends on where
the gate is, and thus determine how big a ``target'' each gate would
present to any quantum control technique designed to generate them.
These calculations show that out of all two-qubit gates, those
locally-equivalent to the B-gate (introduced and described in
\cite{ZVSW2}) present the largest effective targets.

\smallskip

The content of this paper has the following structure.  After a
discussion of the decomposition and parametrisation of $SU(4)$ in
Section \ref{Parametrisation}, we focus on its geometric properties in
Section \ref{Haar_Measure}, where we derive the invariant length
element and Haar measure for the group, presenting the results in both
the original parametrisation and in the context of the representation
of two-qubit gates offered by the local invariants due to Makhlin
\cite{Makhlin}.  We then use this Haar measure to find the volume of
the space of perfect entanglers in Section \ref{PE-sec}.  Section
\ref{QC} gives the invariant volumes of regions surrounding particular
gates of interest, and shows explicitly that these volumes are
entirely dependent on where the gate is located.  The conclusion of
the paper (Section \ref{Conclusions}) is followed by two supplementary
appendices where we review two methods for finding an invariant
measure, the first (A) using the methods of linear algebra and the
second (B) using the properties of metric spaces.

\section{Decomposition and Parametrisation of $SU(4)$}
\label{Parametrisation}
\setcounter{equation}{0}
\renewcommand{\theequation}{\thesection.\arabic{equation}}

All unitary gates operating on two-qubit states are described by a
$4\times 4$ unitary matrix, an element of the compact group $U(4)$.
Any such matrix may be written as an element of $SU(4)$ multiplied by
a complex number of modulus 1, so the sixteen parameters we use to
specify any gate are the phase of this $U(1)$ prefactor (an angle
modulo $\pi/2$) and the fifteen real parameters of $SU(4)$.

Which fifteen parameters we choose is largely up to us; for instance,
we could use the $SU(4)$ polar coordinates \cite{MMM} or the analogues
of the Euler angles familiar from classical mechanics \cite{TBS}.
However, for our purposes, it is much more convenient to utilise the
Cartan decomposition of the Lie algebra of the group (e.g.,
\cite{Helgason,Cahn,KBG,KC}); this allows us to write any element of
$SU(4)$ as a combination of matrices in $SU(2)\otimes SU(2)$ and the
maximal Abelian subgroup $SU(4)\slash SU(2)\otimes SU(2)$ (which
henceforth we will refer to as $\mathcal{A}$ for brevity's sake).

The utility of this decomposition is apparent when we realise that, in
the basis $\{\ket{00},\ket{01},\ket{10},\ket{11}\}$, any operation
that affects only the first qubit is represented by $U_1\otimes I$,
and one affecting only the second is $I\otimes U_2$, where $U_1$ and
$U_2$ are each $2\times 2$ unitary matrices.  These {\em local}
operations, which act separately and independently on the two qubits,
are therefore described by matrices in $SU(2)\otimes SU(2)$.  The
operations that {\em entangle} the two qubits must then be entirely
determined by the matrices from the Abelian subgroup  $\mathcal{A}$.

\bigskip

\begin{figure*}
\centering \includegraphics[scale=0.45]{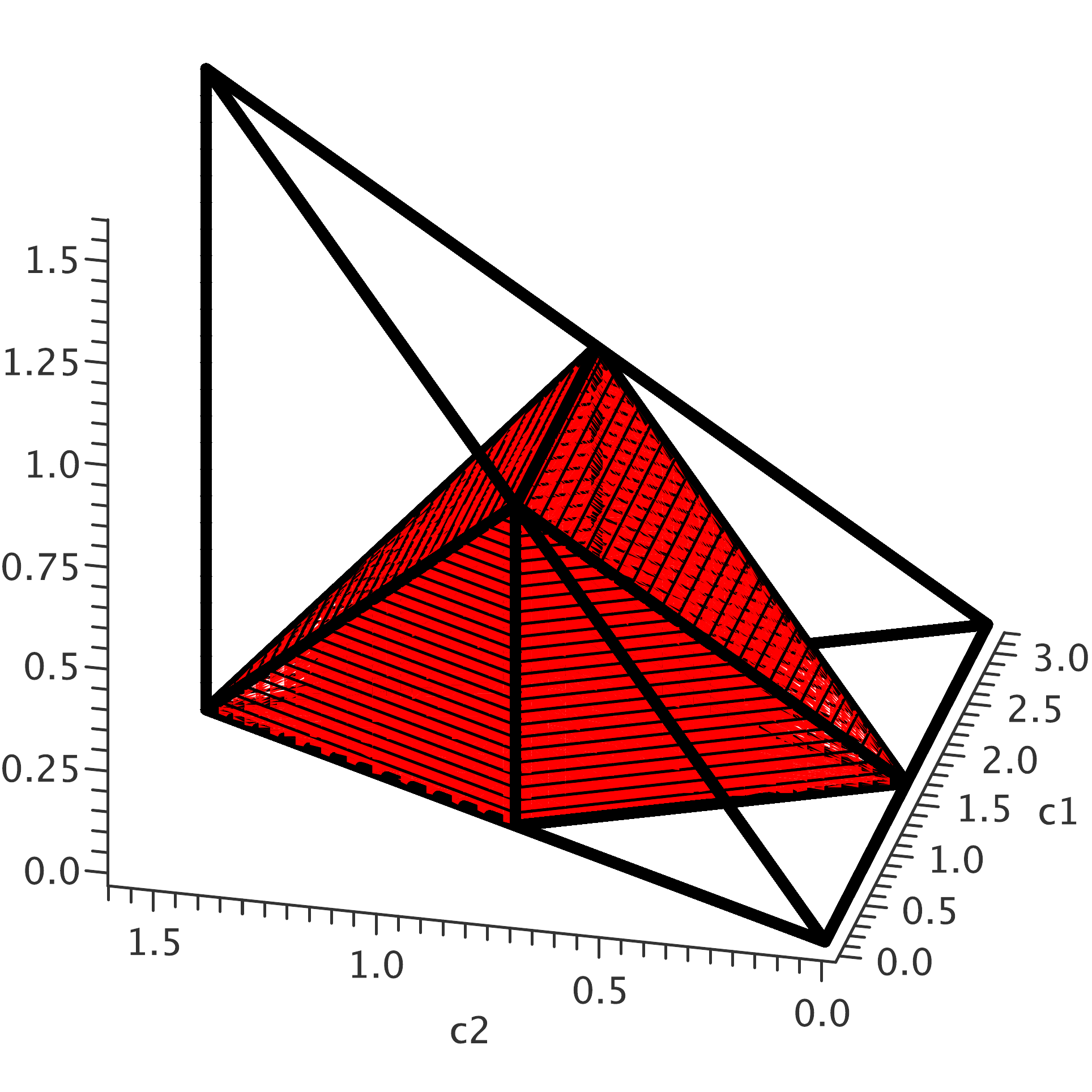}
\caption{(Colour online) The Weyl chamber in $c_1c_2c_3$-space.  The
  perfect entanglers make up the region highlighted in red.}
\label{c-PEs}
\end{figure*}

With all of this in hand, we choose the decomposition of $SU(4)$ such
that our matrices take the form
\begin{eqnarray}
U&=&k_1 A k_2\label{SU(4)},
\end{eqnarray}
where $k_1$ and $k_2$ are $4\times 4$ matrices in $SU(2)\otimes SU(2)$
and $A$ is in the maximal Abelian subgroup $\mathcal{A}$ of $SU(4)$.
We can now parametrise the subgroups in the following way: let
$\vec{\alpha}$ and $\vec{\beta}$ be 3-dimensional vectors given in
terms of spherical coordinates and Cartesian unit vectors by
\begin{eqnarray*}
\vec{\alpha}&=&\alpha\left(\sin\theta\cos\phi
\,\hat{e}_x+
\sin\theta\sin\phi\,\hat{e}_y+\cos\theta\,\hat{e}_z\right)=\alpha
\hat{\alpha},\nonumber\\
\vec{\beta}&=&\beta\left(\sin\lambda\cos\xi\,\hat{e}_x+
\sin\lambda\sin\xi\,\hat{e}_y+\cos\lambda\,\hat{e}_z\right)=\beta\hat{\beta},
\end{eqnarray*}
with $0\leq\alpha,\beta<4\pi$, $0\leq\theta,\lambda<\pi$ and $0\leq
\phi,\xi<2\pi$.  Then if $\sigma_{x,y,z}$ are the usual Pauli
matrices, a generic element of $SU(2)\otimes SU(2)$ may be written as
\begin{eqnarray*}
k\left(\vec{\alpha},\vec{\beta}\right)&=&\exp\left(-\frac{i}{2}
\vec{\alpha}\cdot\vec{\sigma}\right)\otimes\exp\left(-\frac{i}{2}
\vec{\beta}\cdot\vec{\sigma}\right)\nonumber\\
&=&\left[I\cos\left(\frac{\alpha}{2}\right)
-i\hat{\alpha}\cdot\vec{\sigma}
\sin\left(\frac{\alpha}{2}\right)\right]
\otimes\left[I\cos\left(\frac{\beta}{2}\right)
-i\hat{\beta}\cdot\vec{\sigma}
\sin\left(\frac{\beta}{2}\right)\right].
\end{eqnarray*}
The two $SU(2)\otimes SU(2)$ matrices in equation (\ref{SU(4)}) can
then be parametrised by four vectors $\vec{\alpha}_1$,
$\vec{\beta}_1$, $\vec{\alpha}_2$ and $\vec{\beta}_2$ via
\begin{eqnarray*}
k_1=k\left(\vec{\alpha}_1,\vec{\beta}_1\right),&& k_2=k\left(\vec{\alpha}_2,
\vec{\beta}_2\right).
\end{eqnarray*}
This takes care of twelve of the fifteen coordinates necessary to
specify any $SU(4)$ element; the remaining three, $c_1$, $c_2$ and
$c_3$, parametrise the matrix $A$ through
\begin{eqnarray*}
A\left(c_1,c_2,c_3\right)&=&\exp\left(-\frac{i}{2}\sum_{j=1}^3c_j
\sigma_j\otimes\sigma_j\right)\nonumber\\
&=&\prod_{j=1}^3\left[I\otimes I\cos\left(\frac{c_j}{2}\right)-
i\sigma_j\otimes\sigma_j\sin\left(\frac{c_j}{2}\right)\right].
\end{eqnarray*}
To ensure that each $U$ is given by a unique set of coordinates, we
must restrict $c_1$, $c_2$ and $c_3$ to the Weyl chamber given by
\begin{eqnarray*}
0\leq c_3\leq c_2\leq c_1\leq \frac{\pi}{2}
&\mbox{and}&\frac{\pi}{2}<c_1<\pi,\,0\leq c_3\leq c_2<\pi-c_1,
\end{eqnarray*}
i.e., within the tetrahedron whose vertices are at $(0,0,0)$,
$(\pi,0,0)$, $(\pi/2,\pi/2,0)$ and $(\pi/2,\pi/2,\pi/2)$ \cite{ZVSW1},
as shown in Figure \ref{c-PEs}.

Now that we have defined the coordinates and determined their ranges
of values, we can choose an orientation; in this paper, we take the
one such that the ordering
\begin{eqnarray*}
x&=&\left(x^1,\ldots,x^{15}\right)\nonumber\\
&=&\left(\vec{\alpha}_1,\vec{\beta}_1,\vec{\alpha}_2,\vec{\beta}_2,
\vec{c}\right)\nonumber\\
&=&\left(\alpha_1,\theta_1,\phi_1,\beta_1,\lambda_1,\xi_1,
\alpha_2,\theta_2,\phi_2,\beta_2,\lambda_2,\xi_2,c_1,c_2,c_3\right)
\end{eqnarray*}
forms a right-handed coordinate system.

We now want to find a Haar measure for $SU(4)$ in terms of these
fifteen parameters.  The basic method for finding such a measure for
an $N$-dimensional simple compact Lie group $G$ is reviewed in the
appendices, and the first step is to compute the Maurer-Cartan form
$\Theta$ and write it in terms of the $N$ Hermitian Lie algebra
generators $\{T_A\}$ and $N$ coordinate 1-forms $\{\mathrm{d}
x^{\mu}\}$ as
\begin{eqnarray*}
\Theta&=&-iE^A{}_{\mu}(x)T_A\mathrm{d} x^{\mu}.
\end{eqnarray*}
$E$ is therefore a real $N\times N$ matrix whose determinant gives us
our invariant measure (up to an overall factor):
\begin{eqnarray*}
\mathrm{d}\mu&\propto&\abs{\det E(x)}\mathrm{d}^Nx,
\end{eqnarray*}
where $\mathrm{d}^Nx=\mathrm{d}x^1\wedge\ldots\wedge\mathrm{d}x^N$.
Two of the ways of motivating this particular form of the measure are
covered in the appendices, but both require us to somehow compute the
determinant of $E$, which for $SU(4)$ is a $15\times 15$ matrix.

\section{The Invariant Length Element and Haar Measure for
$SU(4)$}\label{Haar_Measure}
\setcounter{equation}{0}
\renewcommand{\theequation}{\thesection.\arabic{equation}}

In this section, we derive expressions for the invariant length
element $\mathrm{d}s^2$ and the Haar measure $\mathrm{d}\mu$ for
$SU(4)$.  Both of these have been found before not just for $SU(4)$,
but for $SU(n)$ and, indeed, for a great variety of simple compact Lie
groups (see, for example \cite{MMM,TBS,SHH} and references therein).
However, the novelty of our approach is that these quantities will be
in forms that are particularly suited for the description of two-qubit
gates, namely, in the coordinate system defined in the previous
section, which separates the purely local gates in $SU(2)\otimes
SU(2)$ from the entangling gates in $\A$.

\subsection{The Length Element}

We choose to do the computation by first finding an invariant length
element $\mathrm{d}s^2$ for $SU(4)$; since this will give the metric
tensor via $\mathrm{d}s^2=g_{\mu\nu}(x)\mathrm{d}x^{\mu}
\otimes\mathrm{d}x^{\nu}$, we may then use the relation $|\det
E|\propto\sqrt{|\det g|}$.  We could also have explicitly found the
full $15\times 15$ matrix $E^A{}_{\mu}$ and then computed its
determinant; this can be done using methods similar to those in
\cite{TBS,SHH}.  However, we found that the computation was somewhat
simpler using $g_{\mu\nu}$ instead; we now describe the calculation
that leads to this.

\smallskip

First, define the three $1$-forms $\Theta_{1,2,\A}$ by
\begin{eqnarray*}
\Theta_1&=&k_1^{-1}\mathrm{d}k_1,\nonumber\\
\Theta_2&=&\mathrm{d}k_2\,k_2^{-1},\nonumber\\
\Theta_{\A}&=&A^{-1}\mathrm{d}A=\mathrm{d}A\,A^{-1}
\end{eqnarray*}
(the latter holding because $\A$ is Abelian).  It is straightforward
to show that the $SU(4)$ Maurer-Cartan form $\Theta$ can be written as
\begin{eqnarray*}
\Theta&=&k_2^{-1}\left(A^{-1}\Theta_1A+\Theta_{\A}+\Theta_2\right)k_2
\end{eqnarray*}
and that the invariant length, given (see Appendix B) by
\begin{eqnarray*}
\mathrm{d}s^2&=&-\mathrm{tr}\left(\Theta\dot{\otimes}\Theta\right),
\end{eqnarray*}
can be expressed as
\begin{eqnarray}
\mathrm{d}s^2&=&-\mathrm{tr}\left(\Theta_1\dot{\otimes}\Theta_1\right)
-\mathrm{tr}\left(\Theta_2\dot{\otimes}\Theta_2\right)
-\mathrm{tr}\left(\Theta_{\A}\dot{\otimes}\Theta_{\A}\right)\nonumber\\
&&-\mathrm{tr}\left(\Theta_1\dot{\otimes}\Theta_{\A}+
\Theta_{\A}\dot{\otimes}\Theta_1\right)
-\mathrm{tr}\left(\Theta_2\dot{\otimes}\Theta_{\A}+
\Theta_{\A}\dot{\otimes}\Theta_2\right)\nonumber\\
&&-\mathrm{tr}\left(A^{-1}\Theta_1A\dot{\otimes}\Theta_2+
\Theta_2\dot{\otimes}A^{-1}\Theta_1A\right).\label{theta-metric}
\end{eqnarray}
The traces can be evaluated quickly if we choose an orthonormal basis
for $SU(4)$; we take the fifteen generators
$T_{0i}=(I\otimes\sigma_i)/2$, $T_{i0}= (\sigma_i\otimes I)/2$ and
$T_{ij}=(\sigma_i\otimes\sigma_j)/2$, $i,j=x,y,z$, which satisfy
\begin{eqnarray*}
\mathrm{tr}\left(T_AT_B\right)&=&\delta_{AB}.
\end{eqnarray*}
$SU(2)\otimes SU(2)$ is spanned by the six matrices
$\{T_{0i},T_{i0}\}$ and $\A$ by the three matrices $\{T_{ii}\}$, so
the matrices $k$ and $A$ are
\begin{eqnarray*}
k\left(\vec{\alpha},\vec{\beta}\right)&=&\exp\left[-i\sum_{j=1}^3
\left(\alpha_jT_{0j}+\beta_jT_{j0}\right)\right],\nonumber\\
A\left(\vec{c}\right)&=&\exp\left[-i\sum_{j=1}^3c_jT_{jj}\right].
\end{eqnarray*}
Using these, we can explicitly compute $\Theta_1$, $\Theta_2$, $A$ and
$\Theta_{\A}$, and thus the length element in (\ref{theta-metric}).
The first three terms give the invariant length elements of
$SU(2)\otimes SU(2)$ (twice) and $\A$, and the next two terms vanish
because the two subspaces are orthogonal to each other.  The remaining
term -- the last -- can be most conveniently written using what we
know about $SU(2)$: the Maurer-Cartan form for this group has the form
\begin{eqnarray*}
\Theta_{SU(2)}&=&e^{i\vec{\alpha}\cdot\vec{\sigma}/2}
\mathrm{d}e^{-i\vec{\alpha}\cdot\vec{\sigma}/2}\nonumber\\
&=&-\frac{i}{2}\sum_i\zeta^i\left(\vec{\alpha}\right)\sigma_i,
\end{eqnarray*}
where the three 1-forms $\zeta^{x,y,z}$ are
\begin{eqnarray*}
\zeta^{x}\left(\vec{\alpha}\right)&=&
\sin\theta\cos\phi\,\mathrm{d}\alpha+2\sin\left(\frac{\alpha}{2}\right)\left[
\sin\left(\frac{\alpha}{2}\right)\sin\phi+ \cos\left(\frac{\alpha}{2}\right)
\cos\theta\cos\phi\right]\mathrm{d}\theta\nonumber\\
&&+2\sin\left(\frac{\alpha}{2}\right)\sin\theta\left[
\sin\left(\frac{\alpha}{2}\right)\cos\theta\cos\phi
-\cos\left(\frac{\alpha}{2}\right)\sin\phi\right]\mathrm{d}\phi,\nonumber\\
\zeta^{y}\left(\vec{\alpha}\right)&=&
\sin\theta\sin\phi\,\mathrm{d}\alpha+2\sin\left(\frac{\alpha}{2}\right)\left[
-\sin\left(\frac{\alpha}{2}\right)\cos\phi+ \cos\left(\frac{\alpha}{2}\right)
\cos\theta\sin\phi\right]\mathrm{d}\theta\nonumber\\
&&+2\sin\left(\frac{\alpha}{2}\right)\sin\theta\left[
\sin\left(\frac{\alpha}{2}\right)\cos\theta\sin\phi
+\cos\left(\frac{\alpha}{2}\right)\cos\phi\right]\mathrm{d}\phi,\nonumber\\
\zeta^{z}\left(\vec{\alpha}\right)&=&
\cos\theta\,\mathrm{d}\alpha-2\sin\left(\frac{\alpha}{2}\right)\cos
\left(\frac{\alpha}{2}\right)\sin\theta\,\mathrm{d}\theta-
2\sin^2\left(\frac{\alpha}{2}\right)\sin^2\theta\,\mathrm{d}\phi.
\end{eqnarray*}
The invariant length element for $SU(4)$ is therefore
\begin{eqnarray}
\mathrm{d}s^2&=&\mathrm{d}s_{SU(2)}^2\left(\vec{\alpha}_1\right)
+\mathrm{d}s_{SU(2)}^2\left(\vec{\beta}_1\right)
+\mathrm{d}s_{SU(2)}^2\left(\vec{\alpha}_2\right)
+\mathrm{d}s_{SU(2)}^2\left(\vec{\beta}_2\right)\nonumber\\
&&+\mathrm{d}c_1\otimes\mathrm{d}c_1+\mathrm{d}c_2\otimes\mathrm{d}c_2
+\mathrm{d}c_3\otimes\mathrm{d}c_3\nonumber\\
&&-\left[\zeta^x\left(\vec{\alpha}_1\right)
\otimes\zeta^x\left(-\vec{\alpha}_2\right)+\zeta^x\left(-\vec{\alpha}_2\right)
\otimes\zeta^x\left(\vec{\alpha}_1\right)\right.\nonumber\\
&&+\left.\zeta^x\left(\vec{\beta}_1\right)
\otimes\zeta^x\left(-\vec{\beta}_2\right)+\zeta^x\left(-\vec{\beta}_2\right)
\otimes\zeta^x\left(\vec{\beta}_1\right)\right]\cos c_2\cos c_3\nonumber\\
&&-\left[\zeta^y\left(\vec{\alpha}_1\right)
\otimes\zeta^y\left(-\vec{\alpha}_2\right)+\zeta^y\left(-\vec{\alpha}_2\right)
\otimes\zeta^y\left(\vec{\alpha}_1\right)\right.\nonumber\\
&&+\left.\zeta^y\left(\vec{\beta}_1\right)
\otimes\zeta^y\left(-\vec{\beta}_2\right)+\zeta^y\left(-\vec{\beta}_2\right)
\otimes\zeta^y\left(\vec{\beta}_1\right)\right]\cos c_1\cos c_3\nonumber\\
&&-\left[\zeta^z\left(\vec{\alpha}_1\right)
\otimes\zeta^z\left(-\vec{\alpha}_2\right)+\zeta^z\left(-\vec{\alpha}_2\right)
\otimes\zeta^z\left(\vec{\alpha}_1\right)\right.\nonumber\\
&&+\left.\zeta^z\left(\vec{\beta}_1\right)
\otimes\zeta^z\left(-\vec{\beta}_2\right)+\zeta^z\left(-\vec{\beta}_2\right)
\otimes\zeta^z\left(\vec{\beta}_1\right)\right]\cos c_1\cos c_2\nonumber\\
&&-\left[\zeta^x\left(\vec{\alpha}_1\right)
\otimes\zeta^x\left(-\vec{\beta}_2\right)+\zeta^x\left(-\vec{\beta}_2\right)
\otimes\zeta^x\left(\vec{\alpha}_1\right)\right.\nonumber\\
&&+\left.\zeta^x\left(\vec{\beta}_1\right)
\otimes\zeta^x\left(-\vec{\alpha}_2\right)+\zeta^x\left(-\vec{\alpha}_2\right)
\otimes\zeta^x\left(\vec{\beta}_1\right)\right]\sin c_2\sin c_3\nonumber\\
&&-\left[\zeta^y\left(\vec{\alpha}_1\right)
\otimes\zeta^y\left(-\vec{\beta}_2\right)+\zeta^y\left(-\vec{\beta}_2\right)
\otimes\zeta^y\left(\vec{\alpha}_1\right)\right.\nonumber\\
&&+\left.\zeta^y\left(\vec{\beta}_1\right)
\otimes\zeta^y\left(-\vec{\alpha}_2\right)+\zeta^y\left(-\vec{\alpha}_2\right)
\otimes\zeta^y\left(\vec{\beta}_1\right)\right]\sin c_1\sin c_3\nonumber\\
&&-\left[\zeta^z\left(\vec{\alpha}_1\right)
\otimes\zeta^z\left(-\vec{\beta}_2\right)+\zeta^z\left(-\vec{\beta}_2\right)
\otimes\zeta^z\left(\vec{\alpha}_1\right)\right.\nonumber\\
&&+\left.\zeta^z\left(\vec{\beta}_1\right)
\otimes\zeta^z\left(-\vec{\alpha}_2\right)+\zeta^z\left(-\vec{\alpha}_2\right)
\otimes\zeta^z\left(\vec{\beta}_1\right)\right]\sin c_1\sin c_2,
\label{SU4-metric}
\end{eqnarray}
where
\begin{eqnarray*}
\mathrm{d}s_{SU(2)}^2\left(\vec{\alpha}\right)&=&
\mathrm{d}\alpha\otimes\mathrm{d}\alpha+4\sin^2\left(\frac{\alpha}{2}\right)
\mathrm{d}\theta\otimes\mathrm{d}\theta+ 4\sin^2\left(\frac{\alpha}{2}
\right)\sin^2\theta\,\mathrm{d}\phi\otimes\mathrm{d}\phi
\end{eqnarray*}
is the $SU(2)$ invariant length element.

\subsection{The Haar Measure}

The metric tensor $g_{\mu\nu}$ can be extracted from
(\ref{SU4-metric}), and, when considered as a $15\times 15$ matrix,
has an associated determinant.  A lengthy but straightforward
calculation gives the result
\begin{eqnarray*}
\det g&=&\left[\vphantom{\sin^2\left(\frac{\beta_2}{2}\right)}
\sin\left(c_1+c_2\right)\sin
\left(c_1-c_2\right)
\sin\left(c_1+c_3\right)\sin\left(c_1-c_3\right)
\sin\left(c_2+c_3\right)\sin\left(c_2-c_3\right)\right.\nonumber\\
&&\left.\times 256\sin^2\left(\frac{\alpha_1}{2}\right)\sin\theta_1
\sin^2\left(\frac{\beta_1}{2}\right)\sin\lambda_1
\sin^2\left(\frac{\alpha_2}{2}\right)\sin\theta_2
\sin^2\left(\frac{\beta_2}{2}\right)\sin\lambda_2\right]^2.
\end{eqnarray*}
Since $\abs{\det E}\propto\sqrt{\abs{\det g}}$, this allows us to
determine, up to a proportionality constant, the Haar measure we want;
to reflect the decomposition of $SU(4)$ into two copies of
$SU(2)\otimes SU(2)$ and $\mathcal{A}=SU(4)\slash SU(2)\otimes SU(2)$,
we write it as
\begin{eqnarray*}
\mathrm{d}\mu&=&\mathrm{d}\mu_{SU(2)}\left(\vec{\alpha}_1\right)\wedge
\mathrm{d}\mu_{SU(2)}\left(\vec{\beta_1}\right)\wedge
\mathrm{d}\mu_{SU(2)}\left(\vec{\alpha_2}\right)\wedge
\mathrm{d}\mu_{SU(2)}\left(\vec{\beta_2}\right)\nonumber\\
&&\wedge\mathrm{d}\mu_{\mathcal{A}}\left(c_1,c_2,c_3\right),
\label{SU4-measure}
\end{eqnarray*}
where $\mathrm{d}\mu_{SU(2)}$ is the normalised $SU(2)$ Haar measure
in spherical coordinates
\begin{eqnarray*}
\mathrm{d}\mu_{SU(2)}(\alpha,\theta,\phi)&=&\frac{1}{8\pi^2}
\sin^2\left(\frac{\alpha}{2}\right)\sin\theta\,\mathrm{d}\alpha\wedge
\mathrm{d}\theta\wedge\mathrm{d}\phi
\end{eqnarray*}
and $\mathrm{d}\mu_{\mathcal{A}}$ is the normalised Haar measure for
the Abelian subgroup given by
\begin{eqnarray*}
\mathrm{d}\mu_{\mathcal{A}}\left(c_1,c_2,c_3\right)&=&\frac{48}{\pi}\left|
\sin\left(c_1+c_2\right)\sin\left(c_1-c_2\right)\sin\left(c_1+c_3\right)
\sin\left(c_1-c_3\right)\right.\nonumber\\
&&\left.\times\sin\left(c_2+c_3\right)\sin\left(c_2-c_3\right)\right|
\mathrm{d}c_1\wedge
\mathrm{d}c_2\wedge\mathrm{d}c_3.
\end{eqnarray*}
(Conveniently, the quantity in the absolute value above is manifestly
nonnegative when $(c_1,c_2,c_3)$ lies in the Weyl chamber, so taking
the absolute value is redundant and we drop it from now on.)  It is
straightforward to confirm that these measures both integrate to unity
over $SU(2)$ and $\A$ respectively.  The normalised Haar measure on
$SU(4)$ is therefore the wedge product of the five measures given:
\begin{eqnarray*}
\mathrm{d}\mu&=&\frac{3}{256\pi^9}\prod_{i=1}^2\left[\sin^2\left(
\frac{\alpha_i}{2}
\right)\sin\theta_i\sin^2\left(\frac{\beta_i}{2}\right)\sin
\lambda_i\right]\nonumber\\
&&\times\prod_{1\leq j<k\leq 3}\left[\sin\left(c_j+c_k\right)
\sin\left(c_j-c_k\right)\right]\,\mathrm{d}^{15} x.
\end{eqnarray*}

Two elements $U$ and $U^{\prime}$ of $SU(4)$ are locally equivalent to
one another if one can be obtained from the other via either left or
right multiplication by an element of $SU(2)\otimes SU(2)$.  In other
words, when $U$ and $U^{\prime}$ are decomposed into the form given in
(\ref{SU(4)}), they have the same matrix $A$.  Thus, any local
equivalence class $[U]\in\A$ is uniquely determined by coordinates
$(c_1,c_2,c_3)$ in the Weyl chamber, and so the invariant measure for
the space of these classes is obtained by integrating over all the
$SU(2)$ parameters.  The result is the normalised Haar measure on
$\A$:
\begin{eqnarray*}
\mathrm{d}\mu_{\mathcal{A}}&=&M_{\mathcal{A}}\left(c_1,c_2,c_3\right)
\mathrm{d}c_1\wedge \mathrm{d}c_2\wedge\mathrm{d}c_3,
\end{eqnarray*}
where
\begin{eqnarray*}
M_{\mathcal{A}}\left(c_1,c_2,c_3\right)&=&
\frac{48}{\pi}\left[\prod_{1\leq j<k\leq 3}\sin\left(c_j+c_k\right)
\sin\left(c_j-c_k\right)\right].
\end{eqnarray*}
Alternatively, using some trigonometric identities and a bit of
algebra, we may rewrite this in a form somewhat more useful for
computations:
\begin{eqnarray}
M_{\mathcal{A}}\left(c_1,c_2,c_3\right)&=&
\frac{3}{\pi}\left[\cos\left(2c_1\right)\cos\left(4c_2\right)
+\cos\left(2c_2\right)\cos\left(4c_3\right)\right.\nonumber\\
&&+\cos\left(2c_3\right)\cos\left(4c_1\right)-\cos\left(4c_1\right)
\cos\left(2c_2\right)\nonumber\\
&&\left.-\cos\left(4c_2\right)\cos\left(2c_3\right)
-\cos\left(4c_3\right)\cos\left(2c_1\right)\right].\label{M-measure}
\end{eqnarray}
As this measure involves only elementary functions, computing the
invariant volume of a region in $\mathcal{A}$ can often be done
exactly, as we will show in Sections \ref{PE-sec} and \ref{QC}.

\subsection{Local Invariants}

We have just derived expressions for the measure and metric in terms
of the three parameters $c_1$, $c_2$ and $c_3$; although both these
expressions are (relatively) simple in form, they are only useful if
we actually have values for these three coordinates.  In practice,
however, extracting $c_1$, $c_2$ and $c_3$ from an arbitrary $SU(4)$
matrix $U$ may be difficult.  Fortunately, there are three far easier
to obtain alternative parameters that can be used as coordinates on
$\mathcal{A}$.

If we change from the standard computational basis
$\{\ket{00},\ket{01},\ket{10},\ket{11}\}$ to the Bell basis
\begin{eqnarray*}
&&\left\{
\frac{1}{\sqrt{2}}\left(\ket{00}-i\ket{11}\right),
-\frac{i}{\sqrt{2}}\left(\ket{01}+\ket{10}\right),
\frac{1}{\sqrt{2}}\left(\ket{01}-\ket{10}\right),
\frac{1}{\sqrt{2}}\left(\ket{00}+i\ket{11}\right)
\right\},
\end{eqnarray*}
then our $SU(4)$ matrices become $U_{\mathrm{B}}=Q^{\dagger}UQ=
Q^{\dagger}k_1Ak_2Q$, where
\begin{eqnarray*}
Q&=&\frac{1}{\sqrt{2}}\left(\begin{array}{cccc}
1&0&0&i\\0&i&1&0\\0&i&-1&0\\1&0&0&-i
\end{array}\right).
\end{eqnarray*}
The eigenvalues of the matrix
$m=U_{\mathrm{B}}^{\mathrm{T}}U_{\mathrm{B}}$ determine all the local
invariants of $U$, also called the Makhlin invariants \cite{Makhlin}.
The characteristic equation of $m$ is
\begin{eqnarray*}
\lambda^4-\tr(m)\lambda^3+\frac{1}{2}\left[\tr^2(m)-\tr\left(m^2\right)
\right]\lambda^2 -\tr^*(m)\lambda+1&=&0
\end{eqnarray*}
and so $\tr(m)$ and $\tr(m^2)$ give local invariants.  These are
complex numbers, so instead we may take as local invariants the three
real numbers
\begin{eqnarray}
&&g_1=\mathrm{Re}\left\{\frac{\tr^2(m)}{16}\right\},\, g_2=\mathrm{Im}
\left\{\frac{\tr^2(m)}{16}\right\},\, g_3=\frac{\tr^2(m)-\tr\left(m^2
\right)}{4}.\label{invariants}
\end{eqnarray}
$m$, $m^2$ and their traces are readily computable using the simplest
of matrix operations, and so values for $g_1$, $g_2$ and $g_3$ can be
easily obtained for any $U\in SU(4)$.

\begin{figure*}
\centering
\includegraphics[scale=0.45]{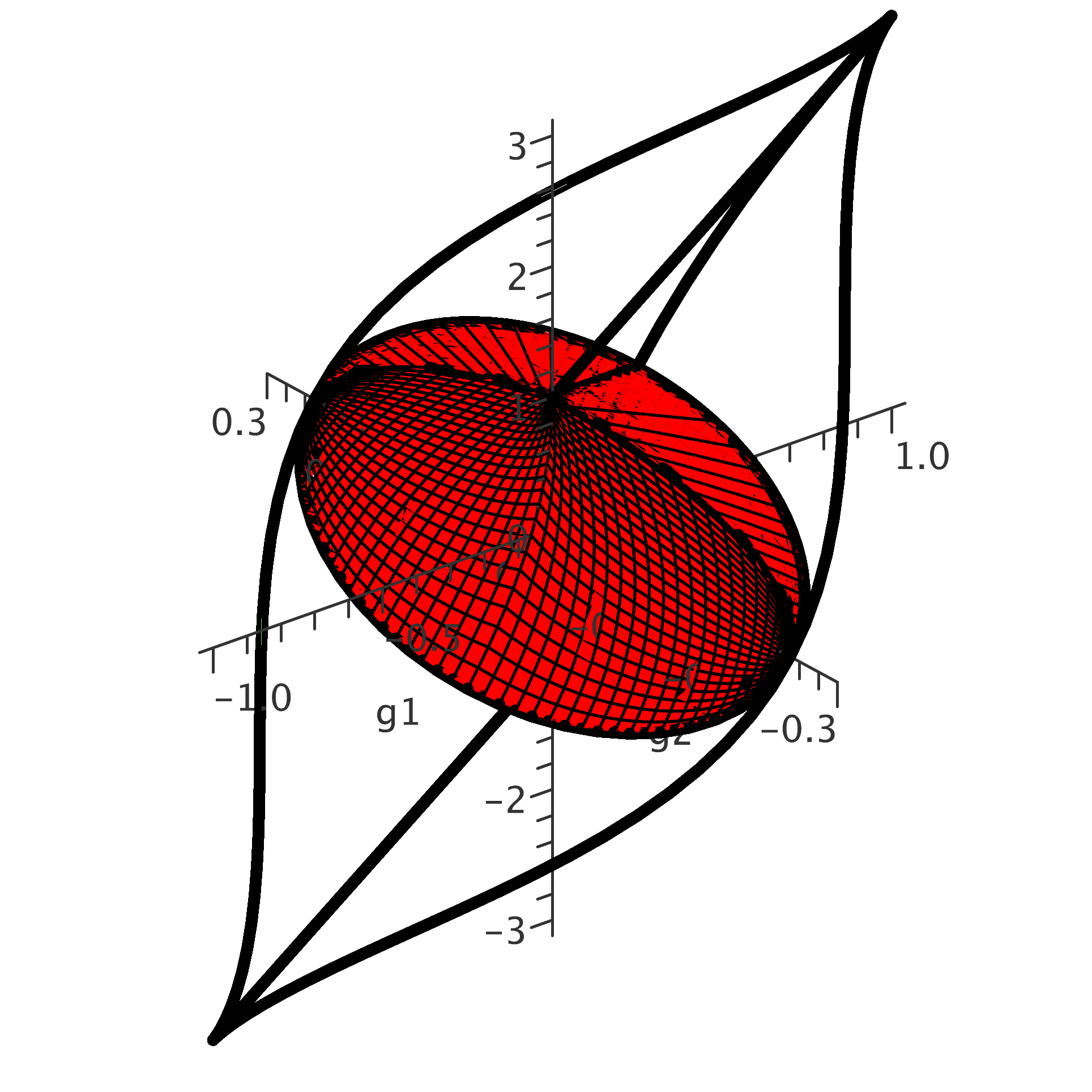}
\caption{(Colour online) The Weyl chamber in $g_1g_2g_3$-space, with
  the region of perfect entanglers highlighted in red.}
\label{EoS}
\end{figure*}

Since these are local invariants, they must be functions only of
$c_1$, $c_2$ and $c_3$; some computation shows that they are, and have
the explicit forms
\begin{eqnarray}
g_1&=&\frac{1}{4}\left[\cos\left(2c_1\right)+\cos\left(2c_2\right)
+\cos\left(2c_3\right)+\cos\left(2c_1\right)\cos\left(2c_2\right)
\cos\left(2c_3\right)\right],\nonumber\\
g_2&=&\frac{1}{4}\sin\left(2c_1\right)\sin\left(2c_2\right)
\sin\left(2c_3\right),\nonumber\\
g_3&=&\cos\left(2c_1\right)+\cos\left(2c_2\right)+\cos\left(2c_3\right).
\label{c-to-g}
\end{eqnarray}
These can be used to embed the Weyl chamber into $g_1g_2g_3$-space.
However, the Weyl chamber is no longer a simple tetrahedron in these
coordinates, but rather an elongated ``Eye of Sauron'' shape
\cite{JRRT,PJ}, as shown in Figure \ref{EoS}.

These functions are bijective when $c_1$, $c_2$ and $c_3$ lie within
the Weyl chamber and we use the following inverse map
$(g_1,g_2,g_3)\mapsto (c_1,c_2,c_3)$: first, find $z_1$, $z_2$ and
$z_3$, the roots of the cubic equation
\begin{eqnarray}
z^3-g_3z^2+\left(4\sqrt{g_1^2+g_2^2}-1\right)z+\left(g_3-4g_1\right)&=&0,
\label{cubic}
\end{eqnarray}
ordered so that $z_1\leq z_2\leq z_3$.  Then $c_2=\cos^{-1}(z_2)/2$,
$c_3=\cos^{-1}(z_3)/2$ and $c_1$ is given by either $\cos^{-1}(z_1)/2$
if $g_2\geq 0$ or $\pi-\cos^{-1}(z_1)/2$ if $g_2<0$.  (As used here,
$\cos^{-1}$ is the principal value of the arccosine function, lying
between $0$ and $\pi$.)

The Haar measure in terms of the local invariants has the relatively
simple form
\begin{eqnarray}
\mathrm{d}\mu_{\mathcal{A}}\left(g_1,g_2,g_3\right)&=&\frac{3}{\pi}
\frac{\mathrm{d}g_1\wedge\mathrm{d}g_2\wedge\mathrm{d}g_3}{
\sqrt{g_1^2+g_2^2}}.
\label{measure-Makhlin}
\end{eqnarray}
However, the form of the length element is much more complicated in
$g_1$, $g_2$ and $g_3$ than it is in $c_1$, $c_2$ and $c_3$: the
Jacobian matrix $J$, which gives the coordinate transformation between
$\vec{c}^{\mathrm{T}}=(c_1,c_2,c_3)$ and
$\vec{g}^{\mathrm{T}}=(g_1,g_2,g_3)$, is defined by
$\mathrm{d}\vec{g}=J\cdot\mathrm{d}\vec{c}$ and has the entries
\begin{eqnarray*}
J_{1i}=-\frac{1}{2}\left[1+\cos\left(2c_j\right)\cos\left(2c_k\right)
\right]\sin\left(2c_i\right),&&j,k\neq i,\, j<k,\nonumber\\
J_{2i}=\frac{1}{2}\cos\left(2c_i\right)\sin\left(2c_j\right)
\sin\left(2c_k\right),&&j,k\neq i,\, j<k,\nonumber\\
J_{3i}=-2\sin\left(2c_i\right).&&
\end{eqnarray*}
The Euclidean length element
$\mathrm{d}c_1^2+\mathrm{d}c_2^2+\mathrm{d}c_3^2$ therefore becomes
$\mathrm{d}\vec{g}^{\mathrm{T}}\cdot(JJ^{\mathrm{T}})^{-1}\cdot\mathrm{d}\vec{g}$,
and this can be written purely in terms of the local invariants:
\begin{eqnarray*}
JJ^{\mathrm{T}}&=&2\left(\begin{array}{ccc}
\rho-4g_1^2+2g_2^2+g_1g_3&g_2g_3-6g_1g_2& 6\rho-2g_1g_3\\
g_2g_3-6g_1g_2&\rho+2g_1^2-4g_2^2-g_1g_3&-2g_2g_3\\
6\rho-2g_1g_3&-2g_2g_3&16\rho+2-2g_3^2
\end{array}\right),
\end{eqnarray*}
where $\rho:=\sqrt{g_1^2+g_2^2}$.  Inverting this matrix is possible
but not particularly illuminating, so we do not do it here.  However,
it illustrates the key feature, that this part of $\mathrm{d}s^2$ can
be written explicitly in terms of the local invariants without needing
to solve (\ref{cubic}).

Unfortunately, the cross-terms in (\ref{SU4-metric}) -- those
involving the $\zeta$-forms -- depend on the local invariants through
$\sin c_i\sin c_j$ and $\cos c_i\cos c_j$, and writing these
explicitly in terms of $g_1$, $g_2$ and $g_3$ leads to an extremely
complicated form for the length element.  Although this part of
$\mathrm{d}s^2$ will not figure into any calculation at a fixed point
in $SU(2)\otimes SU(2)$, if one is to compute the invariant distance
between two arbitrary points in $SU(4)$, it is this form that must be
used if we choose the local invariants as coordinates.

\subsection{Extension to $U(4)$}

We have so far discussed only the two-qubit gates that lie in $SU(4)$
and we will continue to concentrate on this group for the remainder of
this article; however, as stated in the introduction, a general
two-qubit gate will be an element of $U(4)$, so we digress momentarily
to explain how all of the results just obtained may be easily extended
to all of $U(4)$.

This is done through the decomposition $U(4)=[0,\pi/2)\times SU(4)$,
  where the first term in the Cartesian product contributes to an
  overall phase factor:
\begin{eqnarray*}
U&=&e^{i\chi}k_1Ak_2,
\end{eqnarray*}
with $k_1$, $k_2$ and $A$ as before and $\chi\in [0,\pi/2)$
  (considered as a group with addition modulo $\pi/2$).  The invariant
  length element and Haar measure of $U(4)$ are therefore obtained
  from those of $SU(4)$ via, respectively, the addition of
  $4\mathrm{d}\chi\otimes\mathrm{d}\chi$ to (\ref{SU4-metric}) and the
  wedge product of $2\mathrm{d}\chi/\pi$ with (\ref{SU4-measure}).

However, the coordinates $g_1$, $g_2$ and $g_3$ as given in
(\ref{invariants}) will depend on $\chi$, and so must be redefined so
as to be independent of not only the $SU(2)\otimes SU(2)$ local gates,
but also the $U(1)$ phase.  Luckily, this is accomplished by simple
division by the determinant of $U$ \cite{ZVSW1}:
\begin{eqnarray*}
&&g_1=\mathrm{Re}\left\{\frac{\tr^2(m)}{16\det U}\right\},\,
  g_2=\mathrm{Im}\left\{\frac{\tr^2(m)}{16\det U}\right\},\,
  g_3=\frac{\tr^2(m)-\tr\left(m^2\right)}{4\det U}.
\end{eqnarray*}
This modification ensures that the coordinate transformation from
$(c_1,c_2,c_3)$ to $(g_1,g_2,g_3)$ given by (\ref{c-to-g}) remains the
same.  Thus, all our results for $SU(4)$ will easily extend to $U(4)$;
however, for the remainder of this article, we shall once again
concern ourselves only with $SU(4)$.

\section{Perfect Entanglers}\label{PE-sec}
\setcounter{equation}{0}
\renewcommand{\theequation}{\thesection.\arabic{equation}}

The elements of $SU(4)$ that perfectly entangle two-qubit states all
lie within the subset of the Weyl chamber bounded by the planes
$c_1+c_2=\pi/2$, $c_1-c_2=\pi/2$ and $c_2+c_3=\pi/2$.  This region is
the interior of the 7-faced polyhedron with vertices at $(\pi/2,0,0)$,
$(\pi/4,\pi/4,0)$, $(3\pi/4,\pi/4,0)$, $(\pi/2,\pi/2,0)$,
$(\pi/4,\pi/4,\pi/4)$ and $(3\pi/4,\pi/4,\pi/4)$, the red volume
illustrated in Figure \ref{c-PEs}.

At any {\em specific} point in the $SU(2)\otimes SU(2)$ orbit, this
region fills exactly half of the Weyl chamber: if both $k_1$ and $k_2$
are constant, then
$\mathrm{d}s^2=\mathrm{d}c_1^2+\mathrm{d}c_2^2+\mathrm{d}c_3^2$, and
the space is flat.  The Euclidean volume -- calculated with the
normalised measure
$\frac{24}{\pi^3}\,\mathrm{d}c_1\wedge\mathrm{d}c_2\wedge\mathrm{d}c_3$
-- is $1/2$.

However, if we are more concerned with those $SU(4)$ elements that
entangle the two qubits, we are not concerned with what the volume of
the entangling chamber is at a specific point in $SU(2)\otimes SU(2)$;
in fact, since this subgroup only consists of local gates, we are not
interested at all in the values of $k_1$ and $k_2$, but rather only in
those values of $A$ where $(c_1,c_2,c_3)$ is in the
perfectly-entangling chamber.

Therefore, the total volume in $SU(4)$ occupied by the space of
perfect entanglers is obtained by integrating the Haar measure around
the full $SU(2)\otimes SU(2)$ orbit, i.e., all values of
$(\vec{\alpha}_1,\vec{\beta}_1,\vec{\alpha}_2,\vec{\beta}_2)$, as well
as the values of $c_1$, $c_2$ and $c_3$ giving the perfect entanglers.
Since the four $SU(2)$ measures are already normalised, and
$M_{\A}(c_1,c_2,c_3)$ is symmetric around $c_1=\pi/2$, the integral
over the subset of perfect entanglers is
\begin{eqnarray*}
V_{\mathrm{PE}}&=&2\int_{\pi/4}^{\pi/2}\mathrm{d}c_1\left[\int_{\pi/2-c_1}^{\pi/4}
  \mathrm{d}c_2\int_0^{c_2}\mathrm{d}c_3+\int_{\pi/4}^{c_1}\mathrm{d}c_2
  \int_0^{\pi/2-c_2}\mathrm{d}c_3\right]M_{\A}\left(c_1,c_2,c_3\right)\nonumber\\
&=&\frac{8}{3\pi},
\end{eqnarray*}
so we obtain the rather surprising result that the perfect entanglers
occupy over $84\%$ of $SU(4)$!

\smallskip

There are two important remarks to make concerning this result: first,
we chose to do the computation in $c_1c_2c_3$-space because, in these
coordinates, the Haar measure has a relatively simple form and the
boundary of the region of perfect entanglers is bounded by planes,
making the integral of $\mathrm{d}\mu$ very straightforward.  We could
also have chosen to do the integral in $g_1g_2g_3$-space using
(\ref{measure-Makhlin}), but the region of perfect entanglers -- the
red ``pupil'' in Figure \ref{EoS} -- has boundaries much more
complicated than planes, and so the volume integral would be much more
difficult to calculate.  However, the invariance of our measure
ensures that we would obtain the same result of $8/3\pi$ if we did use
the Makhlin invariants.

Secondly, we have shown that perfect entanglers make up a {\em
  majority} of all two-qubit gates.  From the point of view of quantum
information processing, this is good news, because it suggests that it
may be easier than expected to create a perfectly-entangling gate.  In
fact, if we are able to pick a two-qubit gate purely at {\em random},
we would get a perfect entangler nearly 85\% of the time!

It is this second point that we will address in more detail in the
next section: the computation of the invariant volumes of specific
regions in $SU(4)$, those surrounding the types of gates of particular
interest to quantum computing, e.g.,\ the CNOT and SWAP gates.

\smallskip

{\bf Note added in proof:} During the refereeing process following the
submission of this manuscript, we became aware of \cite{MKZ}, in which
two of our results -- the form of the Haar measure on $\A$ and the
volume of the space of perfect entanglers -- were independently
obtained.  However, the technique used in the aforementioned article
differs greatly from ours: the measure was obtained by using results
from the theory of random matrices \cite{RM}, which gives only its
form on $\A$ and not on the entirety of $SU(4)$.  In contrast, our
approach is geometrically motivated and gives much more general
results: we obtain the measure on $\A$ by first constructing an
invariant length element for $SU(4)$ and then using the associated
metric to find a Haar measure for the entire group.  The measure on
$\A$ follows from integration around the orbit of $SU(2)\otimes
SU(2)$.  However, in both cases, once a measure on $\A$ is obtained,
the computation of the volume of the space of perfect entanglers
readily follows.

\section{Uses in Quantum Control}\label{QC}
\setcounter{equation}{0}
\renewcommand{\theequation}{\thesection.\arabic{equation}}

The implementation of any two-qubit quantum computer requires, of
course, quantum gates that operate on the two qubits.  Creating such
gates presents a formidable technical challenge; one must devise a
system in which an element of $SU(4)$ can evolve from an initial state
(most usually the identity element, but in principle any $SU(4)$
matrix) to a final state that is the desired gate.

In practice, however, we cannot create a gate {\em exactly}.  We can
only end up within a certain neighbourhood of a given gate.  For
example, an arbitrary element of $SU(4)$ depends on fifteen parameters
$x^1,\ldots,x^{15}$; if the gate we want is located at the exact point
$(x^{*1},\ldots,x^{*15})$, we will only ever be able to evolve to a
matrix within a certain parameter range around this point, for
example, a cubic region $(x^{*1}\pm\Delta x^1,\ldots,x^{*15}\pm\Delta
x^{15})$.

The likelihood of us being able to evolve the gate into this region
depends on its size: the greater the volume of the region, the bigger
a target it presents for us to shoot at.  Certain gates may be easier
to implement with greater precision if the target volume over a given
parameter range is large; if it is small, then it may be quite
difficult to end up inside the volume, and we may have to increase the
parameter range (and thus lose precision) in order to finish near the
desired gate.

So how do we determine the target sizes?  If $SU(4)$ were a flat
space, then all target sizes would be the same for a given parameter
range; for example, the cubic region described above would have volume
$2^{15}\Delta x^1\ldots\Delta x^{15}$ regardless of what
$(x^{*1},\ldots,x^{*15})$ was.  But we know that $SU(4)$ has a
non-Euclidean metric, and is not flat.  Therefore, the volume of a
region -- obtained by integration of the Haar measure -- can depend on
both the location of the final gate and the range of parameters
describing its neighbourhood.  The resulting volumes will tell us how
large a target the selected gates present for the range of parameters
we choose, and can therefore be used as an indication of how difficult
a gate is to achieve with precision.

\subsection{Volumes of Target Cubes}

As above, we are only concerned with gates that are equivalent up to
local $SU(2)\otimes SU(2)$ operations, so any target volume we compute
will include an integration over all of this subgroup.  Thus, we will
only have to compute integrals over regions of $\A$, since all points
in this Abelian group are indeed distinct modulo local single-qubit
operations.  So if $[U]$ is the equivalence class of the gate $U$, and
$\mathcal{U}$ is a neighbourhood of $[U]$ in $\A$, the volume in
$SU(4)$ that this region occupies is
\begin{eqnarray*}
V(\mathcal{U})&=&\int_{\left(SU(2)\otimes SU(2)\right)
\times\left(SU(2)\otimes SU(2)\right)\times\mathcal{U}}\mathrm{d}\mu
\nonumber\\
&=&\int_{\mathcal{U}}\mathrm{d}\mu_{\A}.
\end{eqnarray*}

The nonzero curvature of $SU(4)$ makes it likely that regions in $\A$
that are described by the same range of coordinates might {\em not}
have the same volumes.  Specifically, if we choose $(c_1,c_2,c_3)$ as
our coordinates in $\A$, a cube of side length $a$ centred at a point
$(c_1^*,c_2^*,c_3^*)$ in the Weyl chamber will not only have a volume
different from $a^3$, but this volume will also vary depending on
where it is centred.

The following results illustrate these properties.  In all cases, the
region integrated over is a cube of side length $a$ centred on the
five basic gates discussed in \cite{MRMYVWCK} (plus two others, for
illustrative purposes) and whose sides are parallel to the $c_1$,
$c_2$ and $c_3$ axes:
\begin{enumerate}
\item $[\mathbb{1}]$ at $(0,0,0)$, with $0\leq a\leq \pi$:
\begin{eqnarray*}
V&=&\frac{3}{2\pi}\left[8a+a\cos(3a)-9a\cos(a)-3\sin(3a)
+12\sin(2a)-15\sin(a)\right].
\end{eqnarray*}
For small $a$, this is $a^9/40\pi+O(a^{11})$.
\item $[\mathrm{SWAP}]$ at $(\pi/2,\pi/2,\pi/2)$, with $0\leq a\leq \pi$:
\begin{eqnarray*}
V&=&\frac{3}{2\pi}\left[8a+a\cos(3a)-9a\cos(a)-3\sin(3a)
+12\sin(2a)-15\sin(a)\right].
\end{eqnarray*}
For small $a$, this is $a^9/40\pi+O(a^{11})$.
\item $[\sqrt{\mathrm{SWAP}}]$ at $(\pi/4,\pi/4,\pi/4)$, with $0\leq
  a\leq\pi/2$:
\begin{eqnarray*}
V&=&\frac{3}{2\pi}\left[2a\sin(3a)+6a\sin(a)+3\cos(3a)
-3\cos(a)\right].
\end{eqnarray*}
For small $a$, this is $8a^6/5\pi+O(a^8)$.
\item $[\mbox{B-gate}]$ at $(\pi/2,\pi/4,0)$, with $0\leq a\leq\pi/4$:
\begin{eqnarray*}
V&=&\frac{3a}{\pi}\left[\cos(a)-\cos(3a)\right].
\end{eqnarray*}
For small $a$, this is $12a^3/\pi+O(a^5)$.
\item $[\mathrm{CNOT}]$/$[\mathrm{CPHASE}]$ at $(\pi/2,0,0)$, with
  $0\leq a\leq\pi/2$:
\begin{eqnarray*}
V&=&\frac{1}{2\pi}\left[8a+7a\cos(3a)-15a\cos(a)
-9\sin(3a)+12\sin(2a)+3\sin(a)\right].
\end{eqnarray*}
For small $a$, this is $4a^5/\pi+O(a^7)$.
\item $[\mathrm{DCNOT}]$ at $(\pi/2,\pi/2,0)$, with $0\leq a\leq\pi/2$:
\begin{eqnarray*}
V&=&\frac{1}{2\pi}\left[8a+7a\cos(3a)-15a\cos(a)
-9\sin(3a)+12\sin(2a)+3\sin(a)\right].
\end{eqnarray*}
For small $a$, this is $4a^5/\pi+O(a^7)$.
\item Gate at $(\frac{\pi}{2},\frac{\pi}{4},\frac{\pi}{4})$, with $0\leq a\leq\pi/4$:
\begin{eqnarray*}
V&=&\frac{1}{2\pi}\left[3\cos{(a)}-3\cos{(3a)}-4a\sin{(3a)}\right].
\end{eqnarray*}
For small $a$, this is $4a^4/\pi+O(a^6)$.
\end{enumerate}
(The upper bounds on the values of $a$ in the above expressions come
from the fact that if the cubes are too big, then we cannot use
equation (\ref{M-measure}), since it is valid only in the Weyl
chamber.  Computing the volumes of larger cubes is possible but
difficult, and we do not do it here.)

The volumes for small values of $a$ are included to provide a means of
comparison: the smaller the cube is, the closer we are to the exact
gate $[U]$, and so if we are to implement this gate with any
reasonable degree of precision, $a$ will have to be small.  The
leading-order term in the small-$a$ expansion therefore gives the
approximate scaling behaviour for each volume, and we see that the
largest volume occurs at the [B-gate] ($V\sim a^3$) and the smallest
at the identity and [SWAP] gates ($V\sim a^9$), with the volumes of
all other gates lying in between.

All controlled gates have equivalence classes that lie on the
$c_1$-axis between the origin and $c_1=\pi/2$, and the
invariant volume of a cube of side length $a$ around each of them can
be computed in the same fashion as the fixed gates above: if the
centre of the cube is at $(c_1^*,0,0)$, then if $0\leq a\leq c_1^*$,
\begin{eqnarray*}
V\left(c_1^*,0,0\right)&=&\frac{1}{2\pi}\left\{8a+a\cos(3a)-9a\cos(a)\right.
\nonumber\\
&&-\left[3a\cos(3a) -3a\cos(a)-3\sin(3a)+9\sin(a)\right]\cos\left(2c_1^*
\right)\nonumber\\
&&\left.+\left[3a\cos(3a)-3a\cos(a)-6\sin(3a)+12\sin(2a)- 6\sin(a)
\right]\cos\left(4c_1^*\right)\right\}\nonumber\\
&=&\frac{a^5}{2\pi}\left\{3-4\cos\left(2c_1^*\right)+\cos\left(4c_1^*\right)
-\frac{a^2}{15}\left[15-26\cos\left(2c_1^*\right)+11\cos\left(4c_1^*\right)
\right]\right.\nonumber\\
&&\left.+\frac{a^4}{5040}\left[819-1640\cos\left(2c_1^*\right)+905
\cos\left(4c_1^*\right)\right]\right\}+O\left(a^{11}\right).
\end{eqnarray*}
Thus, for any $c_1^*>0$, the invariant volume scales as $a^5$.  (For
$c_1^*=\pi/2$, we recover the previous result shared by the [CNOT] and
[CPHASE] gates.)

\smallskip

All of the above gates lie somewhere on the boundary of the Weyl
chamber; if we take a cube of side length $a$ that lies entirely {\em
  within} the Weyl chamber, then its volume as a function of its
centre $(c_1^*,c_2^*,c_3^*)$ is
\begin{eqnarray*}
V(c^*_1,c^*_2,c^*_3)&=&\frac{3a}{2\pi}\sin{(a)}\sin{(2a)}\left[\cos{(2c^*_1)}
\cos{(4c^*_2)}-\cos{(4c^*_1)}\cos{(2c^*_2)}\right.\nonumber\\
&&+\cos{(2c^*_2)}\cos{(4c^*_3)}-\cos{(4c^*_2)}\cos{(2c^*_3)}\nonumber\\
&&\left.+\cos{(4c^*_1)}\cos{(2c^*_3)}-\cos{(2c^*_1)}\cos{(4c^*_3)}\right]
\nonumber\\
&=&\frac{1}{2}a\sin{(a)}\sin{(2a)}M_{\A}\left(c^*_1,c^*_2,c^*_3\right).
\end{eqnarray*}
For small $a$, the prefactor is approximately $a^3$, the Euclidean
volume of the cube, and so in this limit $V/a^3$ is $M_{\A}$, and thus
tells us how much larger or smaller the actual invariant volume is
than the Euclidean volume.

\smallskip

\begin{figure*}
\centering
\includegraphics[scale=0.22]{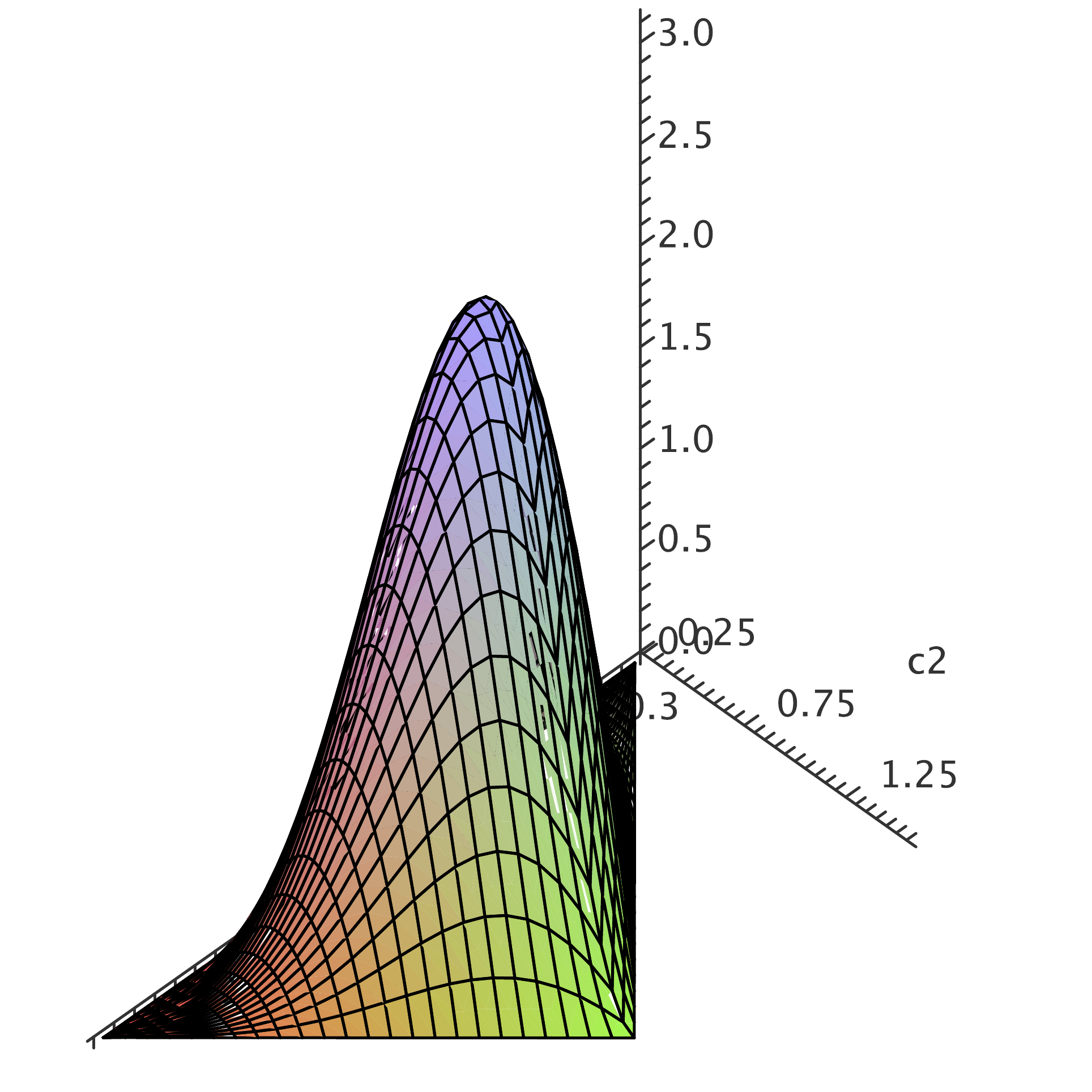}
\includegraphics[scale=0.22]{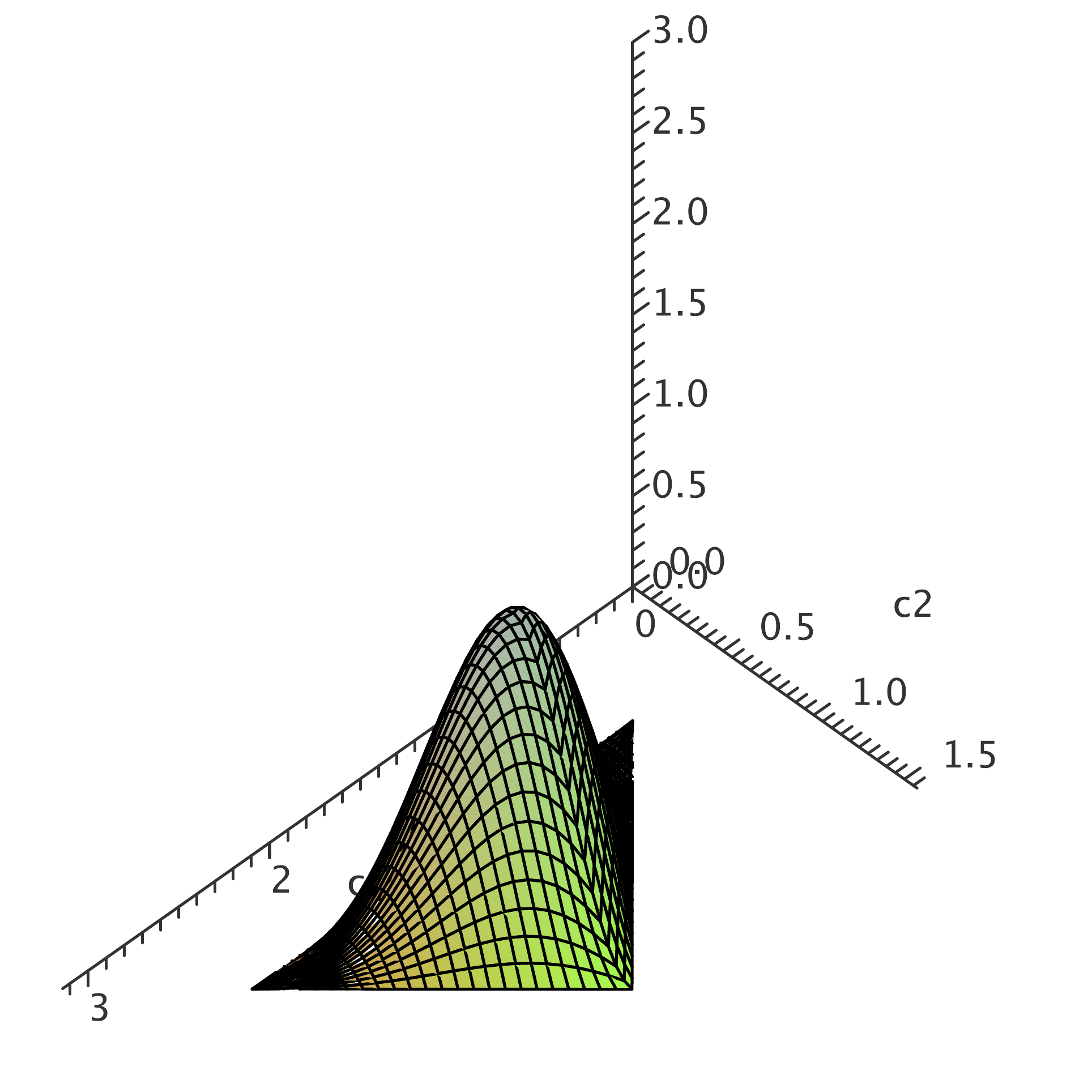}
\includegraphics[scale=0.22]{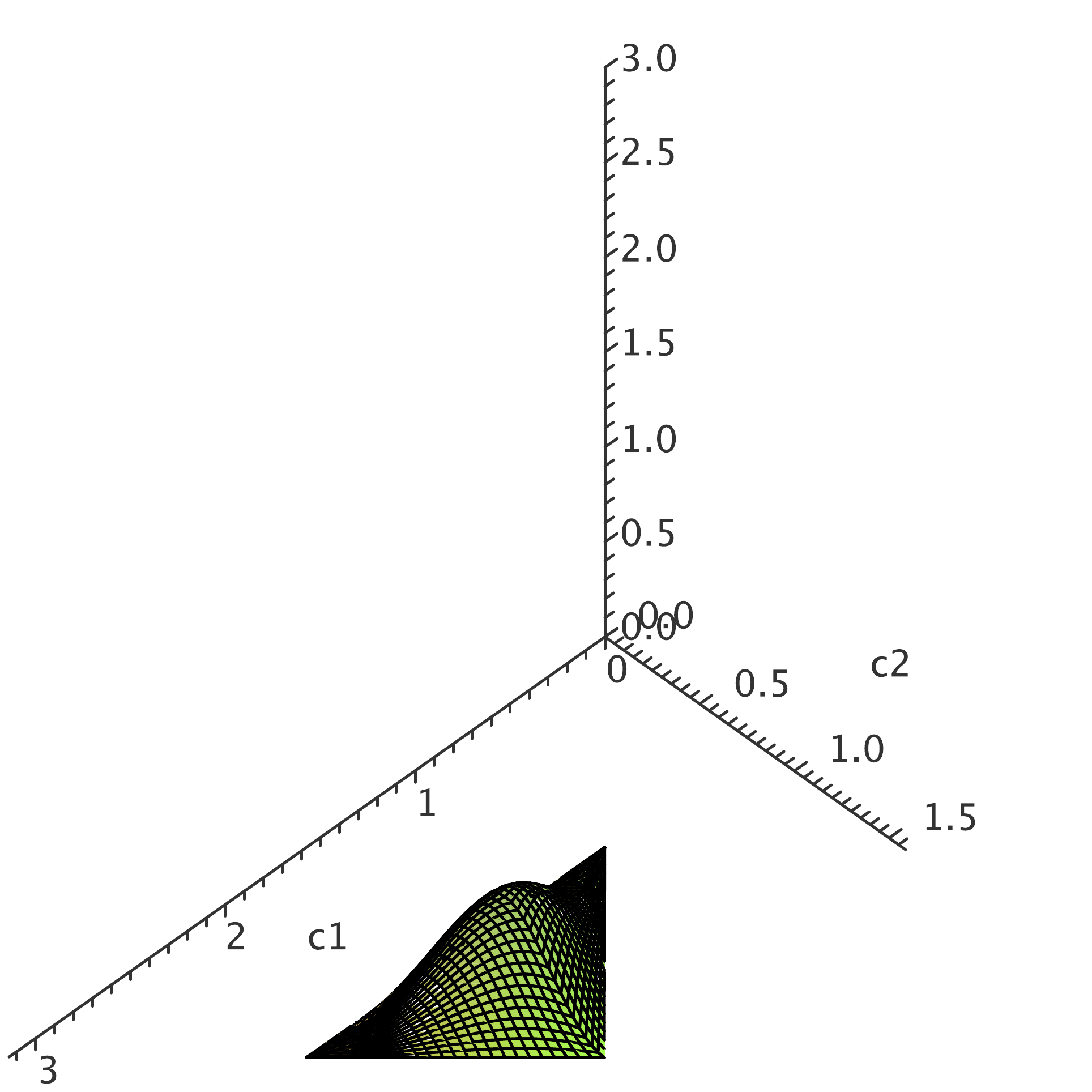}
\caption{(Colour online) Cube volumes within the Weyl chamber.  The
  volume factor $M_{\A}$ as a function of $(c_1^*,c_2^*)$ on
  horizontal slices with, from left to right, $c_3^*=\pi/12$,
  $c_3^*=\pi/6$ and $c_3^*=\pi/4$.}
\label{M-factor}
\end{figure*}

Figure \ref{M-factor} plots $M_{\A}$ for three horizontal slices of
the Weyl chamber, at $c_3^*=\pi/12$, $\pi/6$ and $\pi/4$.  These
illustrate that $M_{\A}$ vanishes on the boundary of the chamber and
peaks in the interior for all $c_3^*>0$.  Furthermore, this maximum
value increases as $c_3^*$ decreases toward zero.  In fact, it is on
this bottom face that $M_{\A}$ takes on its global maximum of $12/\pi$
at $c_1^*=\pi/2$ and $c_2^*=\pi/4$.  This demonstrates that cubes near
the [B-gate] present, for a given side length, the biggest targets.

\subsection{Makhlin Invariants and Target Cylinders}\label{MI}

As is evident from Figure \ref{EoS}, the boundary of the Weyl chamber
in $g_1g_2g_3$-space is no longer a collection of flat planes but a
curved surface.  Computing the volumes of regions that abut the
boundary (precisely where many of the gates of interest are located)
is therefore likely to be far more difficult than in
$c_1c_2c_3$-space.

It is possible, however, to find exact expressions for the volumes of
some regions that lie entirely within the Weyl chamber.  This is most
easily done by converting to cylindrical coordinates $(\rho,\phi,z)$
given by $g_1=\rho\cos\phi$, $g_2=\rho\sin\phi$ and $g_3=z$.  The
measure in these coordinates is very simple:
$3\mathrm{d}\rho\wedge\mathrm{d}\phi\wedge\mathrm{d}z/\pi$.  Using
this, we can explicitly compute the volumes of various regions centred
on the origin:
\begin{eqnarray*}
\mbox{Cube of side length $a$:}&&V=\frac{12a^2}{\pi}\ln\left(\sqrt{2}+1
\right).\nonumber\\
\mbox{Cylinder of height $h$ and axial radius $R$:}&&V=6Rh.\nonumber\\
\mbox{Sphere of radius $R$:}&&V=3\pi R^2.
\end{eqnarray*}

For regions not centred on the origin, the volumes of cubes and
spheres tend to be more difficult to compute, but a closed-form
expression can be found for the volume of a cylinder (with axis in
$g_3$ direction) of height $h$ and radius $R$ centred at
$(g_1^*,g_2^*,g_3^*)$.  If $g_1^*=g_2^*=0$, the volume is the same as
at the origin, namely, $6Rh$.  If either $g_1^*$ or $g_2^*$ is nonzero,
then $\rho^*=\sqrt{(g_1^*)^2+(g_2^*)^2}$ is positive and the invariant
volume of the cylinder is
\begin{eqnarray*}
V\left(g_1^*,g_2^*,g_3^*\right)&=&\left\{
\begin{array}{crl}
\frac{12Rh}{\pi}E\left(\frac{\rho^*}{R}\right)&\mbox{for}&R\geq\rho^*,\\
\frac{12\rho^*h}{\pi}\left[E\left(\frac{R}{\rho^*}\right)+
\left(\frac{R^2}{(\rho^*)^2}-1\right) K\left(\frac{R}{\rho^*}\right)\right]
&\mbox{for}&R<\rho^*,
\end{array}\right.
\end{eqnarray*}
where $K(k)$ and $E(k)$ are the complete elliptic integrals of the
first and second kind respectively:
\begin{eqnarray*}
K(k)=\int_0^{\pi/2}\frac{\mathrm{d}\phi}{\sqrt{1-k^2\sin^2\phi}},&&
E(k)=\int_0^{\pi/2}\mathrm{d}\phi\,\sqrt{1-k^2\sin^2\phi}.
\end{eqnarray*}
For small cylinders with $R\ll\rho^*$, we find
\begin{eqnarray*}
V\left(g_1^*,g_2^*,g_3^*\right)&\approx&\frac{3R^2h}{\sqrt{\left(g_1^*\right)^2
+\left(g_2^*\right)^2}},
\end{eqnarray*}
so the volume of the cylinder decreases as we move away from the
$g_3$-axis, entirely consistent with the result we obtained in
$c_1c_2c_3$-space.

\section{Conclusions} \label{Conclusions}
\setcounter{equation}{0}
\renewcommand{\theequation}{\thesection.\arabic{equation}}

In order to study the geometric properties of $SU(4)$ in a way that is
particularly suitable to a quantum information context -- where the
emphasis is on the entangling capabilities of two-qubit operations --
we have utilised a parametrisation of $SU(4)$ that reflects the
natural decomposition of two-qubit gates into local (single-qubit)
$SU(2)\otimes SU(2)$ and purely nonlocal (two-qubit) $SU(4) /
SU(2)\otimes SU(2)$ factors.  The latter (denoted by $\A$) corresponds
to the maximal Abelian subgroup of $SU(4)$ and is parametrised by
three real coordinates.

In this parametrisation, we have calculated the invariant length
element and the Haar measure of $SU(4)$, with the latter normalised to
provide unit total volume of the group.  These calculations also show
that while the purely nonlocal part of the two-qubit operations is
geometrically flat, the local part carries a curvature that is carried
over to the curvature of $SU(4)$.

We continue with a discussion of the metric properties of the Abelian
subgroup $\mathcal{A}$ of $SU(4)$ in the context of a different choice
of coordinates, namely, the Makhlin invariants.  Although these
invariants are easily determined from a general element of $SU(4)$ and
the Haar measure takes a relatively simple form, the invariant length
element is far more complicated.  Its form can be determined but is not
particularly illuminating; however, the results we present are
sufficient to allow one to compute the invariant distance between two
arbitrary points should the local invariants be selected as the
preferred coordinates for $\A$.

These results allow us to compute the invariant volume of any region
in the Abelian subgroup $\A$ of $SU(4)$, i.e., any region in the space
of local equivalence classes of two-qubit gates.  We first apply it to
the set of perfect entanglers; these gates, which are capable of
creating maximally entangled states out of some product states,
correspond to half of the local equivalence classes.  We found that
the invariant volume of perfect entanglers occupies more than $84\%$
of the total volume of two-qubit gates, which means that, in fact, the
{\em majority} of the two-qubit gates are perfect entanglers.  (Our
form of the Haar measure on $\A$ and our volume of the space of
perfect entanglers are in complete agreement with the recent
independently-obtained results in \cite{MKZ}.)

Next, we use the Haar measure to find the invariant volumes of
locally-equivalent regions around specific gates.  All these regions
are described by the same range of parameters, but due to the
curvature of the space, not all these regions have the same volume.  In
fact, the invariant volumes depend entirely on where in $\A$ the
region lies.  We find that the volume is smallest around the identity
and SWAP gates and largest at the B-gate, with all other volumes
falling in between.

These results are relevant to quantum information processing and its
physical implementation in general, and in particular, to recent
efforts \cite{MRMYVWCK} to use optimal control approach to generate
two-qubit quantum operations, where the control objective is any gate
of a given entangling power rather than a specific two-qubit gate.  In
cases where the objective is to achieve a perfect entangling gate, our
conclusion that the majority of {\em all} gates are perfect entanglers
is highly encouraging.

If the objective is to create one of the more familiar logical gates,
our results show that generating a SWAP gate with any precision may be
difficult due to the low density of gates in its neighbourhood,
whereas the high density near the B-gate suggests that it could be
relatively easy to generate.  Since the B-gate is one of the gates that
is needed to create a universal quantum computer, this is also an
encouraging result.

\section*{Acknowledgements}
\setcounter{equation}{0}
\renewcommand{\theequation}{\thesection.\arabic{equation}}

The authors wish to acknowledge funding from Science Foundation
Ireland under the Principal Investigator Award 10/IN.1/I3013.  We would
also like to thank Mark Howard for his very evocative ``Eye of
Sauron'' description.

\section*{Appendices}
\begin{appendix}

\section{Haar Measures on Compact Lie Groups}

\setcounter{equation}{0}
\renewcommand{\theequation}{\thesection.\arabic{equation}}

Suppose $G$ is a simple compact $N$-dimensional Lie group with
corresponding Lie algebra $g$.  Let $\{ x^{\mu}|\mu=1,\ldots,N\}$ be a
set of local coordinates on the manifold $M$ underlying $G$, with
$\{\mathrm{d} x^{\mu}\}$ the associated $1$-forms.  Given $U(x)\in G$,
we may construct the {\it Maurer-Cartan} $1$-form $\Theta$ as
\begin{eqnarray*}
\Theta&:=&U^{-1}\mathrm{d}U.
\end{eqnarray*}
This $1$-form is left-invariant and right-covariant; in other words,
under the left-translation
\begin{eqnarray*}
U(x)&\mapsto&V U(x),
\end{eqnarray*}
$\Theta$ is unchanged, and under the right-translation
\begin{eqnarray*}
U(x)&\mapsto&U(x) W^{-1},
\end{eqnarray*}
$\Theta$ transforms via conjugation by $W$: $\Theta\mapsto W\Theta W^{-1}$.

We want an invariant measure for $G$, namely, a positive-definite
$N$-form on $M$ that does not change under either the left- or
right-translations above, and thus may play the role of a volume
element on the group.  We construct it by noticing that the wedge
product of $\Theta$ with itself any number of times is also
left-invariant and right-covariant.  Thus, if we have a
finite-dimensional irreducible representation (irrep) $\rho$ of $g$,
then taking the trace of $\Theta^{\wedge N}$ in this irrep returns an
$N$-form that is left-invariant automatically and right-invariant due
to the cyclicity of the trace:
\begin{eqnarray*}
\mathrm{tr}_{\rho}\left(\Theta^{\wedge N}\right)&\mapsto&
\mathrm{tr}_{\rho}\left(W \Theta^{\wedge N}
W^{-1}\right)\nonumber\\ &=&\mathrm{tr}_{\rho}\left(\Theta^{\wedge N}\right).
\end{eqnarray*}
Thus, this is an invariant measure for $G$.  For compact Lie groups,
any such measure is unique up to an overall multiplicative factor, and
is called the {\it Haar measure} $\mathrm{d}\mu$ of the group.

\bigskip

Suppose $\{T_A|A=1,\ldots,N\}$ is a Hermitian basis for the simple
compact Lie algebra $g$.  Since $\Theta$ is a $1$-form that takes
values in $g$, we may write it (using Einstein summation convention)
both in terms of the $1$-forms $\mathrm{d} x^{1,\ldots,N}$ and the
generators $T_{1,\ldots,N}$ as
\begin{eqnarray*}
\Theta&=&-iE^A{}_{\mu}\left( x\right)T_A\mathrm{d} x^{\mu},
\end{eqnarray*}
where each of the $N^2$ components $E^A{}_{\mu}$ is simply a numerical
function of the local coordinates.  If we wedge $\Theta$ with itself
$N$ times, then we obtain
\begin{eqnarray*}
\Theta^{\wedge N}&=&(-i)^NE^{A_1}{}_{\mu_1}\ldots E^{A_N}{}_{\mu_N}
T_{A_1}\ldots T_{A_N}
\mathrm{d} x^{\mu_1}\wedge\ldots\wedge\mathrm{d} x^{\mu_N}\nonumber\\
&=&(-i)^NE^{A_1}{}_{\mu_1}\ldots E^{A_N}{}_{\mu_N}
T_{A_1}\ldots T_{A_N}
\epsilon^{\mu_1\ldots\mu_N}\mathrm{d}^N x,
\end{eqnarray*}
where $\epsilon$ is the $N$-dimensional Levi-Civita symbol and
$\mathrm{d}^N x$ is shorthand for $\mathrm{d}
x^1\wedge\ldots\wedge\mathrm{d} x^N$.  If we think of $E$ as an
$N\times N$ matrix, then
\begin{eqnarray*}
\Theta^{\wedge N}&=&(-i)^N\det E\,\, T_{A_1}\ldots
T_{A_N}\epsilon^{A_1\ldots A_N}\mathrm{d}^N x.
\end{eqnarray*}
We therefore see that
\begin{eqnarray*}
\mathrm{tr}_{\rho}\left(\Theta^{\wedge N}\right)&=&(-i)^N
\mathrm{tr}_{\rho}\left(T_{A_1}\ldots T_{A_N}\epsilon^{A_1\ldots
  A_N}\right)\det E\;\mathrm{d}^N x,
\end{eqnarray*}
where $\rho$ is any irrep of $g$.  The trace is just an overall
multiplicative factor, and since the Haar measure is determined only
up to proportionality, we conclude that
\begin{eqnarray*}
\mathrm{d}\mu&\propto&\left|\det E(x)\right|\mathrm{d}^N x.
\end{eqnarray*}
Taking the absolute value of the determinant ensures that the measure
is positive-definite if the proportionality constant is
positive.  Because $G$ is compact, the integral of this $N$-form over
the underlying manifold $M$ is finite, and so we can fix the constant
of proportionality such that this integral is unity.  This defines the
{\it normalised} Haar measure for a compact simple Lie group:
\begin{eqnarray*}
\mathrm{d}\mu&=&\frac{\left|\det E(x)\right|\mathrm{d}^N x}{\int_M
\left|\det E\left(x'\right)\right|\mathrm{d}^N x'}.
\end{eqnarray*}

\bigskip

An important point: for an arbitrary Lie group $G$, it is possible
that the trace over the generators or the determinant of $E$ could
vanish.  However, both are nonzero if $G$ is simple, which we have
assumed.  But this general method may be extended to nonsimple compact
Lie groups as well: if $G=G_1\times G_2\times\ldots\times G_M$ where
each $G_i$ is compact and simple, then the product of their normalised
Haar measures
\begin{eqnarray*}
\mathrm{d}\mu&=&\mathrm{d}\mu_{G_1}\wedge\mathrm{d}\mu_{G_2}\wedge\ldots
\wedge\mathrm{d}\mu_{G_M}
\end{eqnarray*}
is a positive-definite left- and right-invariant $N$-form, and thus a
normalised Haar measure on $G$.

As an example, consider $U(n)$: this is a nonsimple compact Lie group
that is equal to $[0,2\pi/n)\times SU(n)$, where $[0,2\pi/n)$ is
    considered as a group under addition modulo $2\pi/n$.  Any element
    of $U(n)$ has the form $e^{i\chi}U$, with $\chi\in[0,2\pi/n)$ and
      $U\in SU(n)$.  Then if $\mathrm{d}\mu_{SU(n)}$ is the normalised
      Haar measure for $SU(n)$, then
\begin{eqnarray*}
\mathrm{d}\mu&=&\frac{n\mathrm{d}\chi}{2\pi}\wedge\mathrm{d}
\mu_{SU(n)}
\end{eqnarray*}
is the normalised Haar measure for $U(n)$.
\section{Metric Structures of Simple Lie Groups}

\setcounter{equation}{0}
\renewcommand{\theequation}{\thesection.\arabic{equation}}

Another standard way of obtaining the invariant measure for a compact
Lie group is via the natural metric structure of the underlying
manifold that is induced by the Maurer-Cartan form.  By ``metric
structure'', we mean a way of measuring lengths and distances in the
Lie group: if $x$ and $y$ are the coordinates of the two elements
$U(x)$ and $U(y)$ in $G$, then we want a function $s(x,y)$ that tells
us ``how far'' $U(x)$ and $U(y)$ are from each other.

Since finite lengths can be built up from infinitesimal lengths, we
need a quantity $\mathrm{d}s$ so that the length of a path $\Gamma$
connecting two points is $\int_{\Gamma}\mathrm{d}s$; this is given by
a two-form written in terms of a symmetric metric tensor $g_{\mu\nu}$
via
\begin{eqnarray*}
\mathrm{d}s^2&=&g_{\mu\nu}(x)\mathrm{d}x^{\mu}\otimes\mathrm{d}x^{\nu}.
\end{eqnarray*}
However, we want this length element to be invariant under the action
$U(x)\mapsto V U(x) W^{-1}$, since this gives the coordinate
transformations on $G$.  The Maurer-Cartan form gives us everything we
need to define such an element: define the $N$ Lie algebra-valued
functions $\Theta_1,\ldots,\Theta_N$ as the coefficients of the
coordinate 1-forms, namely,
\begin{eqnarray*}
\Theta&=&\Theta_{\mu}(x)\mathrm{d}x^{\mu}=\left[-iE^A{}_{\mu}(x)T_A\right]
\mathrm{d}x^{\mu}.
\end{eqnarray*}
If we both left- and right-act on $U(x)$ via $V U(x) W^{-1}$, we know
that $\Theta\mapsto W\Theta W^{-1}$; group multiplication only affects
the Lie algebra-valued part of $\Theta$, so
\begin{eqnarray*}
\Theta_{\mu}&\mapsto&W\Theta_{\mu} W^{-1}.
\end{eqnarray*}
Therefore,
\begin{eqnarray*}
\Theta_{\mu}\Theta_{\nu}&\mapsto&W\left(\Theta_{\mu}\Theta_{\nu}\right)W^{-1}.
\end{eqnarray*}
This is neither invariant nor symmetric in $\mu$ and $\nu$; however,
it can be made {\em both} by taking the trace over an irrep $\rho$: in
other words,
\begin{eqnarray*}
g_{\mu\nu}^{(\rho)}&=&-\tr_{\rho}\left(\Theta_{\mu}\Theta_{\nu}\right)
\end{eqnarray*}
satisfies all the properties we need for a metric tensor.  Written in
terms of the generators and the $N\times N$ real matrices $E$, this
becomes
\begin{eqnarray}
g_{\mu\nu}^{(\rho)}&=&\tr_{\rho}\left(T_AT_B\right)E^A{}_{\mu}E^B{}_{\nu}.
\label{metric}
\end{eqnarray}

The trace in the above expression depends on the particular irrep
$\rho$ we use; however, one of the properties of {\em simple} Lie
algebras is that all such traces are proportional to one
another.  Thus, we may simply pick an irrep $\rho_0$ in which to
compute the trace, and all other metrics will differ from it only by
an overall constant of proportionality.  Thus, let $\eta_{AB}$ denote
the trace in equation (\ref{metric}) using $\rho_0$ and let
$g_{\mu\nu}$ be the resulting metric:
\begin{eqnarray*}
g_{\mu\nu}(x)&=&\eta_{AB}E^A{}_{\mu}(x)E^B{}_{\nu}(x).
\end{eqnarray*}
(If we choose the adjoint representation, then $\eta$ is the Killing
metric of the Lie algebra.)  Readers familiar with the Cartan
formalism of general relativity will recognise this; here, $\eta$
plays the role of the (pseudo)Riemannian flat metric and $E$ gives the
components of the vielbein $1$-forms.

\smallskip

We now have a systematic way to compute $\det E$, the function we need
for our invariant measure: first, we note that for simple Lie
algebras, $\eta$ is nonsingular, so
\begin{eqnarray*}
\det g=(\det\eta)(\det E)^2&\Rightarrow&\abs{\det E}\propto\sqrt{\abs{\det
g}}.
\end{eqnarray*}
Second, the invariant measure can be rewritten as
\begin{eqnarray*}
\mathrm{d}s^2&=&g_{\mu\nu}\mathrm{d}x^{\mu}\otimes\mathrm{d}x^{\nu}\nonumber\\
&=&-\tr\left(\Theta_{\mu}\Theta_{\nu}\right)\mathrm{d}x^{\mu}\otimes
\mathrm{d}x^{\nu}\nonumber\\
&=&-\tr\left(\Theta\dot{\otimes}\Theta\right),
\end{eqnarray*}
where the trace is over the chosen irrep $\rho_0$ and $\dot{\otimes}$
denotes both matrix multiplication and tensor product, i.e.,
\begin{eqnarray*}
\rho\left(\Theta\dot{\otimes}\Theta\right)&:=&
\rho\left(\Theta_{\mu}\right)\cdot\rho\left(\Theta_{\nu}\right)\,
\mathrm{d}x^{\mu}\otimes\mathrm{d}x^{\nu}.
\end{eqnarray*}
This formula makes the invariant length extremely straightforward to
compute, and once $g_{\mu\nu}$ is extracted from it, the invariant
measure follows.
\end{appendix}

\end{document}